# Single-crystalline YIG flakes with uniaxial
# in-plane anisotropy and diverse crystallographic orientations

R. Hartmann[1], Seema[1], I. Soldatov[2], M. Lammel[1], D. Lignon[1], X. Y. Ai[1], G. Kiliani[1],

R. Schäfer[2,3], A. Erb[4], R. Gross[4,5], J. Boneberg[1], M. Müller[1], S. T. B. Goennenwein[1],

E. Scheer[1], A. Di Bernardo[1,6,*]

1  Department of Physics, University of Konstanz, 78457 Konstanz, Germany
2  Institute for Emerging Electronic Technologies, Leibinz Institute for Solid State and Materials Science (IFW) Dresden, 01069 Dresden, Germany
3  Institut für Werkstoffwissenschaft, Technische Universität Dresden, D-01062, Dresden, Germany
4  Walther-Meißner-Institut, Bayerische Akademie der Wissenschaften, D-85748 Garching, Germany
5  School of Natural Sciences, Technische Universität München, 85748 Garching, Germany
6  Dipartimento di Fisica "E. R. Caianiello", Università degli Studi di Salerno, I-84084 Fisciano, Italy

*Email: angelo.dibernardo@uni-konstanz.de

## Abstract
We study sub-micron $Y_3Fe_5O_{12}$ (YIG) flakes that we produce via mechanical cleaving and exfoliation of YIG single crystals. By characterizing their structural and magnetic properties, we find that these YIG flakes have surfaces oriented along unusual crystallographic axes and uniaxial in-plane magnetic anisotropy due to their shape, both of which are not commonly available in YIG thin films. These physical properties, combined with the possibility of picking up the YIG flakes and stacking them onto flakes of other van der Waals materials or pre-patterned electrodes or waveguides, open unexplored possibilities for magnonics and for the realization of novel YIG-based heterostructures and spintronic devices.

# I. Introduction

$Y_3Fe_5O_{12}$ (YIG) has become one of the most intensively investigated materials for developing novel spintronic devices. The combination of a high Curie temperature ($T_{Curie}$) of ~ 560 °C in bulk,[1] a Gilbert damping constant ($\alpha$) much lower than that of other magnetic materials (down to $10^{-5}$; Refs 2 and 3), insulating properties due to a large band gap of ~ 2.85 eV at room temperature[4] ($T$), and the possibility of growing it with perpendicular magnetic anisotropy[5-8] (PMA) are all properties that have contributed to great interest in YIG for spintronics and other device-oriented applications.[9] A PMA in combination with a $T_{Curie}$ much higher than room $T$ is desirable, for example, for the realization of spin-transfer-torque (STT) memories or racetrack memories with high density and good thermal stability.[9] The small $\alpha$ of YIG also reduces the switching current needed for STT or enables ultrafast domain wall motion in racetrack memories.[9-10]

YIG belongs to the main materials currently used in magnonics[11-12] because it supports propagation of dissipationless spin-waves over long distances (tens of micrometers at room temperature[13]) thanks to its low $\alpha$. Furthermore, YIG has also been used for the realization of spin-based and microwave electronic components. These components include logic elements[14-15], transistors[16], holographic memories[17], directional couplers[18], multiplexers[19], circulators[20], waveguides[21], filters and resonators[22], generators[23], sensors[24], etc. Also, upon substitutional doping, YIG can also show a large Faraday rotation, which makes it suitable for optical applications[25].

The fabrication of YIG in thin films with close-to-bulk properties by several groups has also led to the synthesis of YIG-based thin film heterostructures and to the realization of both lateral and vertical devices based on them.[26-33] By studying the properties of these devices, it has been shown that YIG can induce magnetism in ultrathin materials with a Dirac-like electronic band structure like graphene or topological insulators transferred or deposited onto YIG, and induce an anomalous Hall effect measurable up to room $T$ (Refs. 30 and 31). It has also been observed that magnon modes in YIG, which can be excited for example by a precession of the YIG magnetization via microwave irradiation, can transport spin angular momentum over long distances (up to tens of micrometers) even at room $T$ (Refs. 32-35). Studies on N/YIG thin film bilayers (N being a non-magnetic metal) have also demonstrated that YIG is a very efficient injector of spin-polarized currents into the N layer, when its magnetization precession is excited.[36-38] Upon excitation of the magnetization precession, the generated spin current can then be detected as a voltage signal in the N, where a spin-to-charge conversion occurs via the inverse spin Hall effect (ISHE) (Refs. 39 and 40). For this process to be efficient, the N layer must have high spin-current to charge-current conversion efficiency[41], which is the reason why Pt is usually used as N. Pt/YIG heterostructures have been widely used in magnonics because they can even amplify the intensity of propagating spin waves. To this purpose, the presence of perpendicular magnetic anisotropy (PMA) seems crucial[42].

Despite the variety of applications for which they are currently studied, YIG thin films also have some intrinsic limitations, which restrict the range of applications of devices based on them. First, YIG thin films exhibit weak in-plane magnetic anisotropy, whereas for some applications in-plane magnetic anisotropy (IMA), particularly uniaxial IMA, would be desirable. Second, there exists only a limited number of commercial substrates with lattice parameters matching those of YIG, onto which YIG can be grown in epitaxial single-crystalline thin film form. This limitation not only restricts the crystallographic orientations of YIG thin films achievable by growth on commercial substrates but also the number of materials that can be grown in single-crystalline form onto YIG for proximity-effect studies. Third, it is challenging to obtain YIG thin films with magnetic properties exactly matching those of bulk YIG because of strain, interdiffusion, and other effects that usually occur at the interfaces between YIG thin films and growth substrates.

Concerning the first limitation, the growth of YIG thin films with uniaxial IMA could be exploited to realize novel logic devices based on N/YIG bilayers, where the ISHE voltage signal can be switched between different states depending on the direction of the applied magnetic field $H$ (parallel or perpendicular) with respect to the YIG magnetic easy axis. Most approaches reported to date to fabricate YIG thin films with uniaxial IMA, which are based on strain engineering or substitutional doping, resulting in a degradation of the magnetic properties of the thin films compared to bulk YIG.[43-46]

Uniaxial IMA would also be useful for heterostructures where YIG is coupled to a superconductor (S) to realize spintronic application with low-energy dissipation, as it has been more systematically reported for other ferromagnetic insulator/S hybrids (FI/S) like, for sample, EuS/Al.[47-48] Since the effects observed scale with the strength of an in-plane uniform magnetic exchange field ($h_{ex}$), which is, in turn, proportional to $T_{Curie}$, YIG is expected to be better as FI material than EuS or EuO ($T_{Curie} \sim 17$ K for EuS[49] and $\sim 69$ K for EuO[50]) for the realization of FI/S hybrids.[51-52]

The second limitation of YIG thin films stems from the size of the cubic unit cell of YIG (lattice constant $a = 12.376$ Å; Ref. 53). This large lattice constant not only restricts the number of available substrates on which lattice-matched single-crystalline YIG thin films can be grown but also hinders the epitaxial growth of a second lattice-matched material on top of the YIG layer. While the possibility of growing YIG thin films with orientation different from the usual (111) orientation[54-55] would help study effects that can be anisotropic with crystal structure like the dispersion relation of magnons in YIG,[56-58] layering another material in epitaxial single-crystalline form on top of YIG would allow for a stronger coupling of such material to YIG.

The third main limitation of YIG thin films stems from their interface with the GGG substrates, which often degrades the magnetic properties of YIG thin films due to strain or interdiffusion (typically of Y from the YIG and Gd or Ga from the GGG). Such interdiffusion, for example, can generate a dead layer at the GGG/YIG interface[59-60] that can lead to a lower saturation magnetization ($M_s$) compared to bulk YIG.[61] Also, due to its paramagnetic behavior, the GGG substrate can negatively affect the dynamic magnetic properties of YIG thin films.[62-65] The recent realization of free-standing YIG nanomembranes, which are detached from the substrates by either chemical or mechanical lift-off,[65-68] is a promising route to overcome these problems. However, the fabrication of YIG nanomembranes is challenging, and their physical properties still need to be systematically optimized to match those of single-crystalline bulk YIG.

Here, we report the fabrication of single-crystalline sub-micron YIG flakes that we realize via mechanical cleaving and exfoliation of YIG single crystals. By studying the crystallographic and magnetic properties of these YIG flakes, which retain the properties of the bulk crystals from which they are obtained, we find that our YIG flakes overcome some of the main limitations of YIG thin films. Our YIG flakes exhibit strong uniaxial IMA due to their shape, they can be produced with diverse crystallographic orientations (with respect to the flake surface), and they are only weakly bound to the substrate on which they are placed, meaning that they are not affected by strain-induced or other detrimental effects from the substrate. In addition, these YIG flakes can be picked up via the dry transfer technique[69] and combined with other single-crystalline materials like van der Waals (vdW) materials to form novel heterostructures and devices for spintronics, also in the superconducting regime.

## II. Experimental

### A. Growth of YIG single crystals

The Pb-free YIG single crystals used in this study have been grown using the traveling solvent floating zone (TSFZ) method. Based on the YIG phase diagram reported in Ref. 70, a composition of 20 mol percent of $Y_2O_3$ and $YFeO_3$ has been used as solvent for the YIG crystal growth using the TSFZ technique. For the growth, a solvent pellet of ~ 0.5 grams has been placed in between the feed and seed rods inside an image furnace. Melting has been obtained at a temperature of ~ 1500 °C in a pure oxygen atmosphere and the molten zone has been moved through the feed rod by moving the mirror system of the image furnace. Thanks to the high solubility of YIG in its flux (~ 50%), a high growth velocity of ~ 4 mm/hour has been achieved during the growth process.

The as-obtained YIG single crystals grow within a few degrees around the [111] crystallographic direction. After growth, the monocrystallinity of the YIG crystals and their orientation have been checked using a Real-Time Laue Back reflection camera (Multiwire Lab Ltd. and Laue-Camera GmbH).

## B. Fabrication of YIG flakes

The fabrication process of YIG flakes starts with cleaving YIG bulk single crystals using a $ZrO_2$ ceramic blade (to prevent contamination of the material) and reducing these crystals into smaller pieces. The $ZrO_2$ blade of ~ 5 cm in length (Carl Roth GmbH manufacturer) is oriented at an angle between 10° and 30° from the direction parallel to the crystal facet to cleave, as shown in Fig. S1a of the Supplementary Material. As a result of the cleaving process, smaller YIG crystals having all lateral dimensions of the order of several hundreds of micrometers are formed. Since the cleaving process is carried out directly on a sticky exfoliation tape (see Figs. S1b and c in the Supplementary Material), the cleaved YIG crystals are ready for the next step consisting in their mechanical exfoliation into the desired YIG flakes with typical thickness between 100 nm and 1000 nm. The exfoliation step can be carried out not only inside a $N_2$ glovebox (as done for other more sensitive van der Waals materials), but also in air because YIG is stable and inert in air. After their mechanical exfoliation, the transfer of YIG flakes onto $SiO_2$(300 nm)/Si substrates with pre-patterned Au/Ti markers is carried out by simply placing the exfoliation tape (with the flakes on top) in contact with the substrate, and slowly peeling off the tape from one corner of the substrate to the opposite one. This results in the transfer of some of the YIG flakes onto the $SiO_2$/Si substrate.

## C. Structural characterization and elemental composition analysis of YIG flakes

The micro-XRD (μ-XRD) measurements for the characterization of the crystallographic structure and orientation of the YIG flakes have been carried out using a Rigaku Smartlab diffractometer. The primary arm of the diffractometer is equipped with a double-bounce channel cut Ge(220) monochromator, which provides a monochromatic CuKα1 (wavelength λ

= 1.5406 Å) radiation. To perform the μ-XRD, the diffractometer has been equipped with a cross-beam optical capillary optics with an incident-limiting slit of 0.5 mm which reduces the beam diameter to ~ 400 μm at the flake position.

Elemental composition analysis of the YIG flakes and confirmation of their crystallographic orientation have been carried out in a scanning electron microscope setup by Energy Dispersive X-ray and Electron Backscatter Diffraction analysis using Oxford Instruments ULTIM MAX and Oxford Instruments SYMMETRY detectors, respectively.

### D. Magneto-optical measurements of YIG flakes

A digitally enhanced wide-field Kerr microscope has been used to investigate the magnetic properties of the YIG flakes.[71] The Kerr microscope has been adjusted for a longitudinal configuration with pure in-plane sensitivity. A blue light generated by light-emitting diodes with a wavelength $\lambda = 457$ nm has been used for the magneto-optical measurements. By sweeping the external magnetic field $H$ along the microscope sensitivity direction and plotting an average grey level of the chosen region of interest as a function of $H$, we have measured the magnetization component $M$ parallel to $H$. A piezo stabilization has also been used to avoid drifting of the image during the measurement process.

## III. Results and discussion

We produce single-crystalline sub-micron-thick flakes of YIG with typical thickness ranging between 100 and 1000 nm using a technique recently developed by our group and already used to produce flakes from other ionic/covalently-bonded materials[72-73] – which allows us to fabricate heterostructures consisting of both vdW and non-vdW flakes.[73] Following the process reported in Sec. II, YIG crystals are mechanically exfoliated into flakes, which are then transferred onto a $SiO_2$ (300 nm)/Si substrate with pre-patterned Au/Ti markers, as shown in Fig. 1(a). As starting YIG single crystals for the above process, we have used YIG single crystals grown in a crucible with Pb flux (Innovent e.V.) as well as other YIG crystals grown using the floating zone method without Pb. The growth process for these YIG single crystals is discussed in Sec. II.

After exfoliation, the substrates are mapped under an optical microscope installed in a glovebox with an inert $N_2$ atmosphere to identify the flakes most likely made of YIG. This is necessary because residues of several other materials are also obtained (mainly from the adhesive tape) as result of the fabrication process. To confirm which flakes, amongst those identified during the mapping of the substrates, are made of YIG, we use energy-dispersive X-ray (EDX) analysis [Fig. 1(b)].

With atomic force microscopy (AFM), we also find that the YIG flakes that we obtain have a thickness typically varying between 100 nm and 1000 nm. Most of them also show a very smooth surface [Fig. 1(b)]. Although the YIG flakes are covalently-bonded like the YIG single crystals from which they are obtained, they can be picked up like flakes of vdW materials and placed onto any vdW flakes via the dry-transfer technique to build hybrid non-vdW/vdW heterostructures. For these applications, YIG flakes with a smooth surface like that shown in Fig. 1(b) are of course desirable.

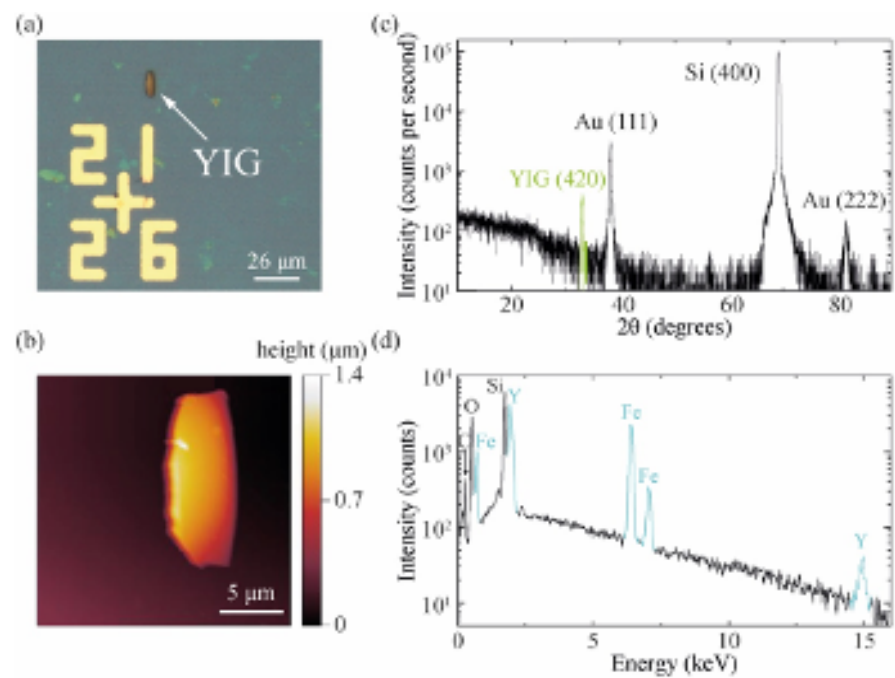

FIG. 1. Structural characterization and elemental composition analysis of sub-micron YIG flakes. (a-b) Optical image of a YIG flake (~1 μm in thickness) on a $SiO_2$(300 nm)/Si substrate with pre-patterned Au/Ti markers (a) and corresponding atomic force microscopy image with surface topography in (b). (c) Micro-XRD high-angle $2\theta/\omega$ scan measured with Cu $K\alpha_1$ radiation of another 850-nm-thick YIG flake (YIG (420) diffraction peak in green) on $SiO_2$/Si substrate with markers (Si and Au diffraction peaks in black) showing that the YIG flake is (210)-oriented. The flake in (b) is (111)-oriented instead. (d) Elemental composition analysis of the flake in (c) based on EDX spectroscopy. The peaks associated with Y and Fe are highlighted in light blue and confirm the stoichiometric 3:5 ratio of Y to Fe.

To determine the crystallographic orientation of our YIG flakes, we use a combination of micro-X-Ray Diffraction (μ-XRD) and Electron Backscatter Diffraction (EBSD) measurements. The high-angle μ-XRD pattern and EBSD analysis [Fig. 1(d) and Fig. 2] show that our YIG flakes not only come with the typical (111) orientation of epitaxial YIG thin films grown on GGG substrates, but we also obtain YIG flakes with different crystallographic orientations. Most of our YIG flakes are in fact (110)-oriented [Fig. 2(b)], whereas others are (210)-, (100)- or (111)-oriented [Figs. 1(d), 2(a) and 2(c)]. We provide statistics about the different orientations of the YIG flakes obtained in the Supplementary Material.

To date, we have not found a specific way to get YIG single crystals that always have a specific crystallographic orientation. This is also the case, when we cleave a specific facet with a well-defined orientation of the original YIG single crystal to get smaller crystals. Also here, we find that, upon further reducing the thickness of the crystals with the exfoliation tape, YIG flakes with a mixture of crystallographic orientations are obtained. We are not sure why this is the case for YIG because for other single crystals like $NiS_2$ reported in previous studies[72], where we have followed the same exfoliation/cleaving process, we usually end up with flakes with a specific orientation, even when different facets of the original crystal are cleaved.

The presence of various crystallographic orientations in our YIG flakes is possibly consistent with the fact that single crystals of garnets like YIG naturally exhibit different facets after growth such as {110}, {210}, {100} facets, in addition to {111} facets.[74-76] We note that the μ-XRD pattern in Fig. 1(c) also shows diffraction peaks from the Au/Ti markers around the flakes due to the size of the beam which has a diameter of ~ 400 μm at the sample position.

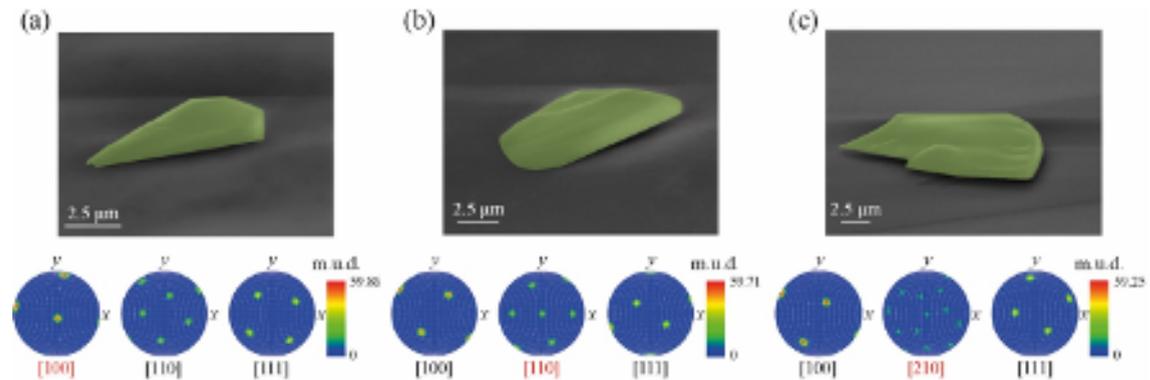

FIG. 2. Crystallographic orientation of YIG flakes. (a-c) Scanning electron micrographs in false color of different YIG flakes (top) with thicknesses of ~ 700 nm, 900 nm and 850 nm and corresponding pole figures determined from EBSD analysis (bottom) along different crystallographic axes (specified below each pole figure). The color bars for the pole figures are given in multiples of uniform density (m.u.d.) units, while the axis perpendicular to the flake surface (determined from the analysis of the pole figures) is highlighted in red below the corresponding pole figure.

To determine whether our YIG flakes exhibit uniaxial IMA, which is usually absent in epitaxial YIG thin films, we characterize the magnetic properties of the flakes by magneto-optical magnetometry at room $T$. For these measurements, we select YIG flakes with an ellipsoidal shape, for which we would expect a magnetic easy axis coinciding with the long axis of the flake, if shape anisotropy were the dominant contribution to magnetic anisotropy.

The results of the magneto-optical magnetometry measurements that we have carried out on the (111)-oriented elongated YIG flake in Figs. 1(a)-1(b) are reported in Fig. 3 for three different orientations of the applied external magnetic field $H$ with respect to the flake's long axis ($a$). We have chosen this flake with a relatively large thickness of ~ 1 μm to increase the intensity of the signal above the resolution of our setup.

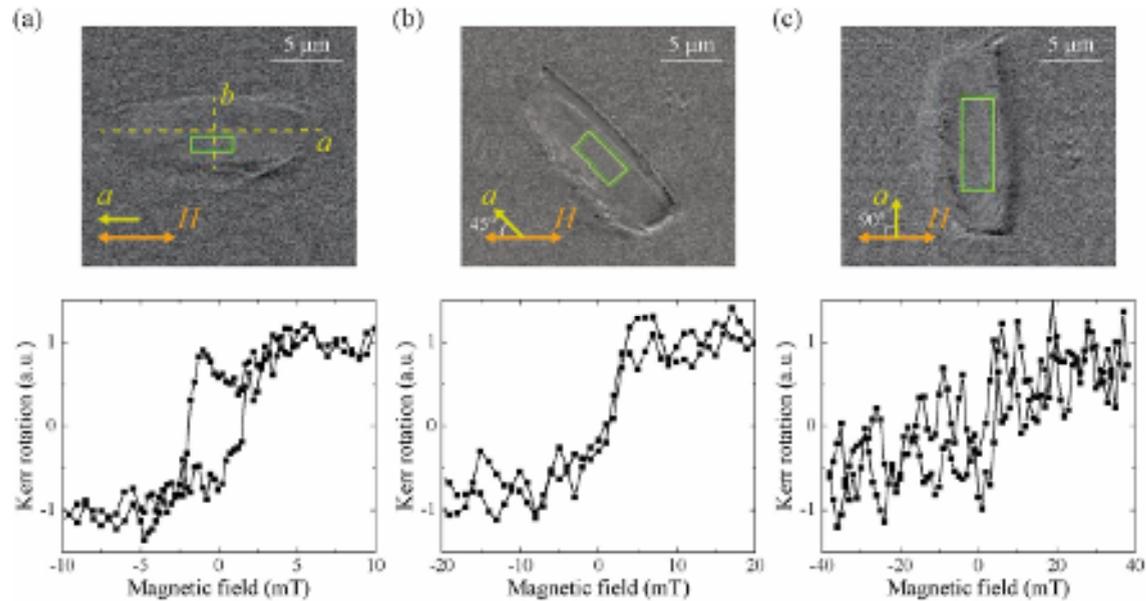

FIG. 3. Magnetic characterization of YIG flake. (a-c) Magneto-optical magnetometry on a (111)-oriented elongated YIG flake (top panels) with thickness ~1 μm for different orientations of the applied magnetic field $H$ with respect to the long axis $a$ of the flake (orientation specified in the left corner of each panel) with corresponding variation of the intensity of the Faraday rotation measured with longitudinal sensitivity as a function of $H$ (bottom panels).

We note here that YIG is almost transparent in the visible region of the electromagnetic spectrum,[77-78] meaning that it does not show a Kerr effect, since Kerr rotation requires measuring absorption.[79-80] YIG, however, can be studied using the magneto-optical Faraday effect, which is a transmission-based effect. The Faraday effect can be resolved by optical transmission polarization microscopy or by placing a YIG sample on top of a non-magnetic mirror in a reflection microscope. In this configuration, the plane-polarized light goes through the sample twice, doubling the intensity of the Faraday rotation (the Faraday rotation is irreversible). Since our YIG flakes are placed on a SiO$_2$/Si substrate, the substrate acts as a mirror with the bottom interface of the YIG flake, leading to an increase in the Faraday rotation signal.

For the magneto-optical magnetometry measurements, we illuminate our YIG flake with blue light (wavelength $\lambda$ = 457 nm) and used a microscope adjusted for pure in-plane sensitivity.[71] The external $H$ during the measurements has been applied parallel to the sensitivity direction, i.e., within the plane of the YIG flake in Fig. 3. By plotting the average grey level of a chosen region of interest as a function of the applied $H$, we can measure the magnetization component $M$ parallel to $H$.

The data in Fig. 3(a) show that the magnetic hysteresis loop, $M(H)$, has a pronounced squareness when $H$ is applied along $a$. Upon rotation of the flake, as $H$ gets progressively misaligned with respect to $a$ [Figs. 3(b) and 3(c)], a reduction in the squareness of the hysteresis loop is observed together with a decrease in the coercive field ($H_c$) from ~ 2.5 mT to 0 mT, which is consistent with what is expected for shaped-induced uniaxial IMA. Correspondingly, the saturation field ($H_s$) of the YIG flake increases from 5 mT for $H \parallel a$ [Fig. 3(a)] to higher values (~ 30 mT) for $H \perp a$ [Fig. 3(c)]. As shown in the Supplementary Material, the measured values of $H_c$ and $H_s$ fit well to those calculated using a Stoner-Wohlfarth approach under the assumption of dominant shape anisotropy in a magnetic ellipsoid with the same dimensions as the YIG flake in Fig. 3 (for the calculations, we took 15 μm x 5 μm in lateral size and 1 μm in thickness). Our magneto-optical measurements therefore suggest that our YIG flakes can have dominant shape anisotropy, which for an ellipsoidal flake like that shown in Fig. 3 results in a magnetic easy axis coinciding with the long axis of the flake.

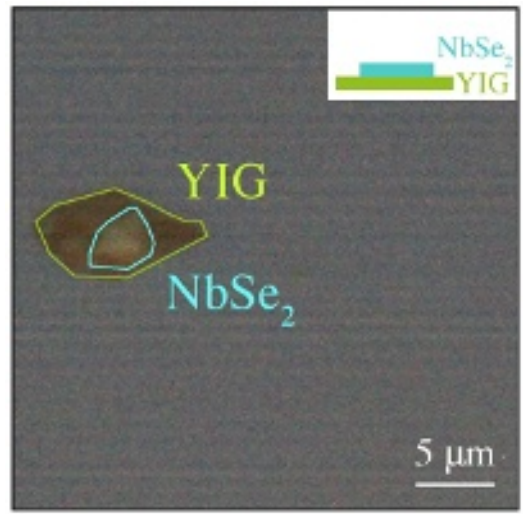

FIG. 4. Coupling of YIG to other vdW materials. Optical microscope image of a heterostructure fabricated by the dry-transfer method and consisting of a flake of a vdW superconductor (NbSe$_2$) stacked onto a YIG flake (the inset shows the materials stack from top to bottom).

We have also carried out magneto-optical magnetometry in polar configuration (i.e., with $H$ and microscope sensitivity both out-of-plane) on the same flake shown in Fig. 3 to determine whether any out-of-plane magnetization reversal occurs. Since no changes in the light polarization rotation signal have been observed, we conclude that the magnetization of our YIG flake in Fig. 3 has no out-of-plane component.

We note that our YIG flakes are entirely detached from their substrate and can be picked up and transferred as typically done for flakes of other vdW materials. Figure 4 shows an example of heterostructures consisting of a YIG flake that we have picked up from its $SiO_2$/Si substrate after fabrication and placed onto another clean substrate using the dry-transfer technique. Before this process, if a YIG flake with a specific orientation must be transferred, then this flake must be pre-selected by performing EBSD measurements on several flakes until the desired orientation is obtained.

With the same dry-transfer technique, a second nanoflake of a vdW superconductor ($NbSe_2$) has been then stacked onto YIG to form the $NbSe_2$/YIG heterostructure in Fig. 4. This example shows that our YIG flakes can be used to fabricate novel material hybrids consisting of sub-micron YIG flakes coupled to other vdW materials. In addition, the heterostructure in Fig. 4 suggests that our YIG flakes can also be placed onto pre-patterned electrodes or devices (e.g., waveguides) or onto transparent substrates to perform a variety of magnetotransport, ferromagnetic resonance or optical transmission experiments, which usually require several fabrication and patterning steps to be carried out on YIG-based thin film heterostructures.

## IV. Conclusions

In conclusion, we have fabricated YIG flakes by cleaving and subsequent mechanical exfoliation of YIG single crystals and characterized their structural and magnetic properties at room $T$. Our analysis shows that the YIG flakes obtained are single-crystalline and exhibit surfaces oriented along different crystallographic axes, most of which are difficult to get in single-crystalline YIG thin films. Also, unlike YIG thin films, our YIG flakes with elongated shape exhibit strong uniaxial in-plane magnetic anisotropy that it is not obtained by strain or doping.

Being able to fabricate sub-micron YIG flakes featuring various crystallographic orientations can pave the way for studies aiming at investigating how the magnon dispersion relation in our YIG flakes varies depending on crystallographic orientation[81,82], and determine whether other orientations lead to a shift in the excitation frequency or in longer propagation lengths for magnonic excitations in YIG. In addition, since our YIG flakes are confined in their lateral dimensions, they are naturally suitable to track the propagation of magnonic excitations optically, without any need for patterning. All these studies can have a significant impact on the development of YIG-based magnonics.

Another significant advantage of our YIG flakes is that they can be picked up via the same dry-transfer technique used for vdW materials. As a result, these YIG flakes can be transferred onto pre-patterned arrays of electrodes to do lateral transport experiments or be embedded in other nanoscale devices like waveguides to perform experiments under FMR excitations, but in the absence of any extrinsic contributions due to the substrate and using YIG material with bulk properties, since our sub-micron flakes are obtained directly from YIG single crystals.

In addition to the above, YIG flakes can be picked up and stacked onto other vdW materials with different properties (topological insulators, superconductors, and normal metals) to study novel exotic phases emerging from their combination or make new spintronic devices. In particular, the in-plane uniaxial anisotropy of our YIG flakes can be exploited to make room-temperature spin valves with very large magnetoresistance or to induce a strong reversible modulation of the superconducting state in an ultrathin vdW superconductor sandwiched between two YIG flakes.

## SUPPLEMENTARY MATERIAL

See the supplementary material for further details on the fabrication of YIG flakes and on their topography characterization, for statistics on the orientation of the YIG flakes, for a calculation of the magnetic anisotropy of YIG flakes based on the Stoner-Wohlfarth model.

## ACKNOWLDEGMENTS

R. H., E. S. and A. D. B. acknowledge funding from the Alexander von Humboldt Foundation in the framework of a Sofja Kovalevskaja grant. A. D. B. also thanks the University of Konstanz for support through a Zukunftskolleg Research Fellowship and, together with E. S., acknowledges funding from the Deutsche Forschungsgemeinschaft (DFG) through the SPP 2244 priority program (grant No. 443404566). S., X. Y. A., M. L., M. M., S. T. B. G. and E. S. also thank the DFG for support through the SFB 1432 (grant No. 425217212). S. also acknowledges support from the University of Konstanz through RiSC funding (Blue Sky Research).

## AUTHOR DECLARATIONS

### Conflict of interest

The authors declare no conflicts of interest.

### Author contributions

**Roman Hartmann**: Investigation (equal); Data curation (equal); Writing – original draft (supporting). **Seema**: Investigation (supporting); Data curation (supporting). **Ivan Soldatov**: Investigation (supporting); Data curation (supporting); Writing – review and editing (supporting). **Michaela Lammel**: Investigation (supporting); Writing – review and editing (supporting). **Daphné Lignon**: Investigation (supporting). **Xian Yue Ai**: Investigation (supporting). **Gillian Kiliani**: Investigation (supporting). **Rudolf Schäfer**: Investigation (supporting); Resources (supporting); Writing – review and editing (supporting). **Andreas Erb**: Investigation (supporting); Resources (supporting); Writing – review and editing (supporting). **Rudolf Gross**: Resources (supporting); Writing – review and editing (supporting). **Johannes Boneberg**: Resources (supporting); Writing – review and editing (supporting). **Martina Müller**: Resources (supporting); Writing – review and editing (supporting). **Sebastian Goennenwein**: Supervision (supporting); Resources (supporting); Writing – review and editing (supporting); **Elke Scheer**: Supervision (equal); Resources (supporting); Funding acquisition (supporting); Writing – review and editing (supporting). **Angelo Di Bernardo**: Conceptualization (lead); Data curation (equal); Supervision (equal); Funding acquisition (equal); Resources (equal); Writing – original draft, review, and editing (lead).

## DATA AVAILABILITY

Data will be made available from the corresponding authors upon request.

# REFERENCES

1. E. E. Anderson, *Phys. Rev. Lett*. 134, A1581 (1964). https://doi.org/10.1103/PhysRev.134.A1581
2. H. Chang, P. Li, W. Zhang, T. Liu, A. Hoffmann, L. Deng, and M. Wu, *IEEE Magn. Lett*. 5, 6700104 (2014). https://doi.org/10.1109/LMAG.2014.2350958
3. Q. B. Liu, K. K. Meng, Z. D. Xu, T. Zhu, X. G. Xu, J. Miao, and Y. Jiang, *Phys. Rev. B* 101, 174431 (2020). https://doi.org/10.1103/PhysRevB.101.174431
4. R. Metselaar, and P. K. Larsen, *Solid State Commun*. 15, 291-294 (1974). https://doi.org/10.1016/0038-1098(74)90760-1
5. M. Kubota, K. Shibuya, Y. Tokunaga, F. Kagawa, A. Tsukazaki, Y. Tokura, and M. Kawasaki, *J. Magn. Magn. Mater.* 339, 63-70 (2013). https://doi.org/10.1016/j.jmmm.2013.02.045
6. J. Fu, M. Hua, X. Wen, M. Xue, S. Ding, M. Wang, P. Yu, S. Liu, J. Han, C. Wang, H. Du, Y. Yang, and J. Yang, *Appl. Phys. Lett*. 110, 202403 (2017). https://doi.org/10.1063/1.4983783
7. H. Wang, C. Du, P. C. Hammel, and F. Yang, *Phys. Rev. B* 89, 134404 (2014). http://dx.doi.org/10.1103/PhysRevB.89.134404
8. G. Li, H. Bai, J. Su, Z. Z. Zhu, Y. Zhang, and J. W. Cai, *APL Mater*. 7, 041104 (2019). https://doi.org/10.1063/1.5090292
9. J. Ding, C. Liu, Y. Zhang, U. Erugu, Z. Quan, R. Yu, E. McCollum, S. Mo, Y. Yang, H. Ding, X. Xu, J. Tang, X. Yang, and M. Wu, *Phys. Rev. Appl*. 14, 014017 (2020). https://doi.org/10.1103/PhysRevApplied.14.014017
10. S. S. P. Parkin, M. Hayashi, and L. Thomas, *Science* 320, 190-194 (2008). https://doi.org/10.1126/science.1145799
11. A. A. Serga, A. V. Chumak, and. B. Hillebrands, *J. Phys. D: Appl. Phys*. 43, 264002 (2010). https://doi.org/10.1088/0022-3727/43/26/264002
12. A. V. Chumak, V.I. Vasyuchka, A.A. Serga, and B. Hillebrands, *Nat. Phys*. 11 (2015) 453-461. https://doi.org/10.1038/nphys3347.
13. H. Qin, R. B. Hollander, L. Flajsman, and S. van Dijken, *Nano Lett.* 22, 5294 (2022). https://doi.org/10.1021/acs.nanolett.2c01238
14. A. Khitun, M. Bao, and K. L. Wang, *J. Phys. D. Appl. Phys*. 43, 264005 (2010). https://doi.org/10.1088/0022-3727/43/26/264005.
15. T. Schneider, A. A. Serga, B. Leven, B. Hillebrands, R. L. Stamps, and M. P. Kostylev, *Appl. Phys. Lett*. 92, 022505 (2008). https://doi.org/10.1063/1.2834714
16. A. V. Chumak, A.A. Serga, and B. Hillebrands, *Nat. Commun*. 5, 4700 (2014). https://doi.org/10.1038/ncomms5700.
17. F. Gertz, A. Kozhevnikov, Y. Filimonov, and A. Khitun, *IEEE Trans. Magn*. 51, 4002905 (2015). https://doi.org/10.1109/TMAG.2014.2362723.
18. Q. Wang, M. Kewenig, M. Schneider, R. Verba, F. Kohl, B. Heinz, M. Geilen, M. Mohseni, B. Lägel, F. Ciubotaru, C. Adelmann, M. Dubs, S.D. Cotofana, O. V. Dobrovolskiy, T. Brächer, P. Pirro, A. V. Chumak, *Nat. Electron*. 3, 765-774 (2020). https://doi.org/10.1038/s41928-020-00485-6.
19. C. S. Davies, A. V. Sadovnikov, S. V. Grishin, Yu. P. Sharaevsky, S. A. Nikitov, and V. V. Kruglyak, *IEEE Trans. Magn*. 51, 3401904 (2015). https://doi.org/10.1109/TMAG.2015.2447010
20. K. Szulc, P. Graczyk, M. Mruczkiewicz, G. Gubbiotti, and M. Krawczyk, *Phys. Rev. Appl*. 14, 034063 (2020). https://doi.org/10.1103/PhysRevApplied.14.034063.
21. Y. V. Khivintsev, V.K. Sakharov, A. V. Kozhevnikov, G.M. Dudko, Y.A. Filimonov, and A. Khitun, *J. Magn. Magn. Mater*. 545, 168754 (2022). https://doi.org/10.1016/j.jmmm.2021.168754.
22. J. Krupka, B. Salski, P. Kopyt, and W. Gwarek, *Sci. Rep*. 6, 34739 (2016). https://doi.org/10.1038/srep34739
23. S. L. Vysotskii, A. V. Sadovnikov, G.M. Dudko, A. V. Kozhevnikov, Y. V. Khivintsev, V. K. Sakharov, N. N. Novitskii, A. I. Stognij, and Y. A. Filimonov, *Appl. Phys. Lett*. 117, 102403 (2020). https://doi.org/10.1063/5.0018388.
24. M. Inoue, A. Baryshev, H. Takagi, P.B. Lim, K. Hatafuku, J. Noda, and K. Togo, *Appl. Phys. Lett*. 98, 132511 (2011). https://doi.org/10.1063/1.3567940.
25. M. Huang, and S. Zhang, *Appl. Phys. A* 74, 177-180 (2002). https://doi.org/10.1007/s003390100883

26. C. Dubs, O. Surzhenko, R. Thomas, J. Osten, T. Scheider, K. Lenz, J. Grenzer, R. Hübner, and E. Wendler, *Phys. Rev. Mater*. 4, 024416 (2020).
https://doi.org/10.1103/PhysRevMaterials.4.024416

27. B. Bhoi, B. Kim, Y. Kim, M.-K. Kim, J.-H. Lee, and S.-K. Kim, *J. Appl. Phys*. 123, 203902 (2018).
https://doi.org/10.1063/1.5031198

28. L. Jin, K. Jia, Y. He, G. Wang, Z. Zhong, and H. Zhang, *Appl. Surf. Sci*. 483, 947-952 (2019).
https://doi.org/10.1016/j.apsusc.2019.04.050

29. T. Shang, Q. F. Zhan, H. L. Yang, Z. H. Zuo, Y. L. Xie, L. P. Liu, S. L. Zhang, Y. Zhang, H. H. Li, B. M. Wang, Y. H. Wu, S. Zhang, and R.-W. Li, *Appl. Phys. Lett*. 109, 032410 (2016)
https://doi.org/10.1063/1.4959573

30. Z. Jiang, C.-Z. Chang, C. Tang, P. Wei, J. S. Moodera, and J. Shi, *Nano Lett*. 15, 5835-5840 (2015).
https://doi.org/10.1021/acs.nanolett.5b01905

31. Z. Wang, C. Tang, R. Sachs, Y. Barlas, and J. Shi, *Phys. Rev. Lett*. 114, 016603 (2015).
https://doi.org/10.1103/PhysRevLett.114.016603

32. L. J. Cornelissen, J. Liu, R. A. Duine, J. B. Youssef, and B. J. van Wees, *Nat. Phys*. 11, 1022-1026 (2015). https://doi.org/10.1038/nphys3465

33. M. Althammer, S. Meyer, H. Nakayama, M. Schreier, S. Altmannshofer, M. Weiler, H. Huebl, S. Geprägs, M. Opel, R. Gross, D. Meier, C. Klewe, T. Kuschel, J.-M. Schmalhorst, G. Reiss, L. Shen, A. Gupta, Y.-T. Chen, G. E. W. Bauer, E. Saitoh, and S. T. B. Gonnenwein, *Phys. Rev. B* 87, 224401 (2013). https://doi.org/10.1103/PhysRevB.87.224401

34. S. T. B. Gonnenwein, R. Schlitz, M. Pernpeintner, K. Ganzhorn, M. Althammer, R. Gross, and H. Huebl, *Appl. Phys. Lett*. 107, 172405 (2015). https://doi.org/10.1063/1.4935074

35. J. Li, Y. Xu, M. Aldosary, C. Tang, Z. Lin, S. Zhang, R. Lake, and J. Shi, *Nat. Commun*. 7, 10858 (2016). https://doi.org/10.1038/ncomms10858

36. Y. Kajiwara, K. Harii, S. Takahashi, J. Ohe, K. Uchida, M. Mizuguchi, H. Umezawa, H. Kawai, K. Ando, K. Takanashi, S. Maekawa, and E. Saitoh, *Nature* 464, 262-266 (2010).
https://doi.org/10.1038/nature08876

37. K.-I. Uchida, T. An, Y. Kajiwara, M. Toda, and E. Saitoh, *Appl. Phys. Lett*. 99, 212501 (2011).
https://doi.org/10.1063/1.3662032

38. M. Haertinger, C. H. Back, J. Lotze, M. Weiler, S. Geprägs, H. Huebl, S. T. B. Goennewein, and G. Woltersdorf, *Phys. Rev. B* 92, 054437 (2015). https://doi.org/10.1103/PhysRevB.92.054437

39. E. Saitoh, M. Ueda, H. Miyajima, and G. Tatara, *Appl. Phys. Lett*. 88, 182509 (2006).
https://doi.org/10.1063/1.2199473

40. J. Sinova, S. O. Valenzuela, J. Wunderlich, C. H. Back, and T. Jungwirth, *Rev. Mod. Phys*. 87, 1213-1259 (2015). https://doi.org/10.1103/RevModPhys.87.1213

41. Y.V. Nikulin, Y.V. Khivintsev, M.E. Seleznev, S.L. Vysotskii, A.V. Kozhevnikov, V.K. Sakharov, G.M. Dudko, A. Khitun, S.A. Nikitov, and Y.A. Filimonov (2023), pre-print available at https://arxiv.org/abs/2311.15096

42. H. Merbouche, B. Divinskiy, D. Gouéré, R. Lebrun, V. Cros, and P. Bortolotti (2023), pre-print available at https://doi.org/10.48550/arXiv.2303.04695.

43. P. C. Van, S. Surabhi, V. Dongquoc, R. Kuchi, S.-G. Yoon, and J.-R. Jeong, *Appl. Surf. Sci*. 435, 377-383 (2018). https://doi.org/10.1016/j.apsusc.2017.11.129

44. O. Galstyan, H. Lee, S. Lee, N. Yoo, J. Park, A. Babajanyan, B. Friedman, and K. Lee, *J. Magn. Magn. Mater*. 366, 24-27 (2014). https://doi.org/10.1016/j.jmmm.2014.04.045

45. E. R. Rosenberg, K. Litzius, J. M. Shaw, G. A. Riley, G. S. D. Beach, H. T. Nembach, and C. A. Ross, *Adv. Electron. Mater*. 7, 2100452 (2021). https://doi.org/10.1002/aelm.202100452

46. J. Mendil, M. Trassin, Q. Bu, M. Fiebig, and P. Gambardella, *Appl. Phys. Lett*. 114, 172404 (2019).
https://doi.org/10.1063/1.5090205

47. J. S. Moodera, X. Hao, G. A. Gibson, and R. Meservey, *Phys. Rev. Lett*. 61, 637-640 (1988).
https://doi.org/10.1103/PhysRevLett.61.637

48. E. Strambini, V. N. Golovach, G. De Simoni, J. S. Moodera, F. S. Bergeret, and F. Giazotto, *Phys. Rev. Mater.* 1, 054402 (2017). https://doi.org/10.1103/PhysRevMaterials.1.054402

49. P. Wachter, in *Handbook on the Physics and Chemistry of Rare Earths*, edited by K. A. Gschneider, Jr., and L. Eyring (North-Holland, Amsterdam, 1979), Chap. 19.

50. M. Müller, G.-X. Miao, and J. S. Moodera, *J. Appl. Phys*. 105, 07C917 (2009).
https://doi.org/10.1063/1.3063673

51. K.-R. Jeon, J.-C. Jeon, X. Zhou, A. Migliorini, J. Yoon, and S. P. Parkin, *ACS Nano* 14, 15874-15883 (2020). https://doi.org/10.1021/acsnano.0c07187

52. K.-R. Jeon, J.-K. Kim, J. Yoon, J.-C. Jeon, H. Han, A. Cottet, T. Kontos, and S. P. Parkin, *Nat. Mater.* 21, 1008-1013 (2022). https://doi.org/10.1038/s41563-022-01300-7

53. S. Geller, *Z. fur Krist.* 125, 1-47 (1967). https://doi.org/10.1524/zkri.1967.125.16.1

54. A. Mitra, O. Cespedes, Q. Ramasse, M. Ali, S. Marmion, M. Ward, R. M. D. Brydson, C. J. Kinane, J. F. K. Cooper, S. Langridge, and B. J. Hickez, *Sci. Rep.* 7, 11774 (2017). https://doi.org/10.1038/s41598-017-10281-6

55. Y. Yang, T. Liu, L. Bi, and L. Deng, *J. Alloys Compd.* 860, 158235 (2021). https://doi.org/10.1016/j.jallcom.2020.158235

56. S. Neusser, and D. Grundler, *Adv. Mater.* 21, 2927-2932 (2009). https://doi.org/10.1002/adma.200900809

57. B. Lenk, H. Ulrichs, F. Garbs, and M. Münzenberg, *Phys. Rep.* 507, 107-136 (2011). https://doi.org/10.1016/j.physrep.2011.06.003

58. N. Träger, F. Groß, J. Förster, K. Baumgaertl, H. Stoll, M. Weigand, G. Schütz, D. Grundler, and J. Gräfe, *Sci. Rep.* 10, 18146 (2020). https://doi.org/10.1038/s41598-020-74785-4

59. J. F. K. Cooper, C. J. Kinane, S. Langridge, M. Ali, B. J. Hickey, T. Niizeki, K. Uchida, E. Saitoh, H. Ambaye, and A. Glavic, *Phys. Rev. B* 96, 104404 (2017). https://doi.org/10.1103/PhysRevB.96.104404

60. S. M. Suturin, A. M. Korovin, V. E. Bursian, L. V. Lutsev, V. Bourobina, N. L. Yakovlev, M. Montecchi, L. Pasquali, V. Ukleev, A. Vorobiev, A. Devishvili, and N. S. Sokolov, *Phys. Rev. Mater.* 2, 104404 (2018). https://doi.org/10.1103/PhysRevMaterials.2.104404

61. G. Schmidt, C. Hauser, P. Trempler, M. Paleschke, and E. Th. Papaloannu, *Phys. Status Solidi B* 257, 1900644 (2020). https://doi.org/10.1002/pssb.201900644

62. H. Huebl, C. W. Zollitsch, J. Lotze, F. Hocke, M. Greifenstein, A. Marx, R. Gross, and S. T. B. Goennenwein, *Phys. Rev. Lett.* 111, 127003 (2013). https://doi.org/10.1103/PhysRevLett.111.127003

63. Y. Tabuchi, S. Ishino, T. Ishikawa, R. Yamazaki, K. Usami, and Y. Nakamura, *Phys. Rev. Lett.* 113, 083603 (2014). https://doi.org/10.1103/PhysRevLett.113.083603

64. P. Trempler, R. Dreyer, P. Geyer, C. Hauser, G. Woltersdorf, and G. Schmidt, *Appl. Phys. Lett.* 117, 232401 (2020). https://doi.org/10.1063/5.0026120

65. V. V. Danilov, D. L. Lyfar, Yu. V. Lyubon'ko, A. Yu. Nechiporuk, and S. M. Ryabchenko, *Sov. Phys. JETP* 32, 276-280 (1989). https://doi.org/10.1007/BF00897267

66. H. S. Kum. H. Lee, S. Kim, S. Lindemann, W. Kong, K. Qiao, P. Chen, J. Irwin, J. H. Lee, S. Xie, S. Subramanian, J. Shim, S.-H. Bae, C. Choi, L. Ranno, S. Seo, S. Lee, J. Bauer, H. Li, K. Lee, J. A. Robinson, C. A. Ross, D. G. Schlom, M. S. Rzchowski, C.-B. Eom, and J. Kim, *Nature* 578, 75-81 (2020). https://doi.org/10.1038/s41586-020-1939-z

67. L. Zhang, D. Zhang, L. Jin, B. Liu, H. Meng, X. Tang, M. Li, S. Liu, Z. Zhong, and H. Zhang, *APL Mater.* 9, 061105 (2021). https://doi.org/10.1063/5.0054595

68. F. Heyroth, C. Hauser, P. Trempler, P. Geyer, F. Syrowatka, R. Dreyer, S. G. Ebbinghaus, G. Woltersdorf, and G. Schmidt, *Phys. Rev. Appl.* 12, 054031 (2019). https://doi.org/10.1103/PhysRevApplied.12.054031

69. R. Frienda, E. Navarro-Moratalla, P. Gant, D. Perez De Lara, P. Jarillo-Herrero, R. V. Gorbachev, and A. Castellanos-Gomez, *Chem. Soc. Rev.* 47, 53-68 (2018). https://doi.org/10.1039/C7CS00556C

70. H. J. van Hook, *J. Am. Ceram. Soc.* 45, 162-165 (1962). https://doi.org/10.1111/j.1151-2916.1962.tb1112.x

71. I. V. Soldatov, and R. Schäfer, *Rev. Sci. Instrum.* 88, 073701 (2017). https://doi.org/10.1063/1.4991820

72. R. Hartmann, M. Hogen, D. Lignon, A. K. C. Tan, M. Amado, S. El-Khatib, M. Egilmez, B. Das, M. Leighton, M. Atature, E. Scheer, and A. Di Bernardo, *Nanoscale* 15, 10277 (2023). https://doi.org/10.1039/D3NR00467H

73. A. Spuri, D. Nikolić, S. Chakraborty, M. Klang, H. Alpern, O. Millo, H. Steinberg, W. Belzig, E. Scheer, and A. Di Bernardo, pre-print at https://arxiv.org/abs/2305.02216

74. V.A. Timofeeva and N.I. Lukyanova, in: Growth of Crystals, Vol. 9, Eds, N.N. Sheftal and E.I. Givargizov (Consultants Bureau, New York, 1975).

75. F. Rinne, and L. Kulaszewski, *Tschermaks Mineral. Pettog. Mitt*. 38, 376-381 (1925). https://doi.org/10.1007/BF02993941

76. E. Beregi, E. Sterk, F. Tanos, E. Hartmann, and J. Lábár, *J. Crys. Growth* 65, 562-567 (1983). https://doi.org/10.1016/0022-0248(83)90103-3

77. F. J. Kahn, P. S. Pershan, and J. P. Remeika, *Phys. Rev.* 186, 891-918 (1969). https://doi.org/10.1103/PhysRev.186.891

78. J. F. Dillon, Jr., *J. Appl. Phys*. 29, 539-541 (1958). https://doi.org/10.1063/1.1723215

79. W. Kuch, R. Schäfer, P. Fischer, and F. U. Hillebrecht, *Magnetic Microscopy of Layered Structures* (Springer-Verlag, Berlin, 2015).

80. A. Hubert, and R. Schäfer, *Magnetic Domains* (Springer, Berlin, 1998).

81. Y.E. Kuzovlev, N.I. Mezin, and Y. V Medvedev, *Low Temp. Phys.* 40, 915-921 (2014). https://doi.org/10.1063/1.4892644

82. R.G. Kryshtal, and A. V. Medved, *J. Magn. Magn. Mater*. 426, 666-669 (2017). https://doi.org/10.1016/j.jmmm.2016.10.148.